\documentclass[useAMS,usenatbib]{mn2e}
\usepackage{amsmath, amssymb, amsfonts}
\usepackage{graphicx}
\usepackage{multicol}
\usepackage{lipsum}
\usepackage{tikz}
\usepackage{float}
\usepackage{color}

\usepackage{rotating}
\usepackage{setspace}
\usepackage{caption}
\usepackage{subcaption}
\usepackage{multirow}
\usepackage{footnote}
\usepackage{titlesec}
\usepackage{hyperref}
\hypersetup{colorlinks=true, citecolor=black, linkcolor=blue, menucolor=black}  

\titlespacing{\section}{0pt}{\parskip}{-\parskip}
\titlespacing{\subsection}{0pt}{\parskip}{-\parskip}
\titlespacing{\subsubsection}{0pt}{\parskip}{-\parskip}

\renewcommand{\thefootnote}{\roman{footnote}}

\title[New detections of HC$_{5}$N toward hot cores]{New detections of HC$_{5}$N toward hot cores associated with 6.7 GHz methanol masers}
\author[Green et al. 2013]{C.-E. Green$^{1,2\star}$,  J. A. Green$^{2,3}$, M. G. Burton$^{1}$, S. Horiuchi$^{4}$, \newauthor N. F. H. Tothill$^{5}$, A. J. Walsh$^{6}$, C. R. Purcell$^{7}$, J. E. J. Lovell$^{8}$, T. J. Millar$^{9}$ \\ 
\\
$^{1}$School of Physics, University of New South Wales, Sydney, NSW, 2052, Australia \\
$^{2}$CSIRO Astronomy \& Space Science, Australia Telescope National Facility, PO Box 76, Epping, NSW 2121, Australia \\
$^{3}$SKA Organisation, Jodrell Bank Observatory, Lower Withington, Macclesfield SK11 9DL, UK \\
$^{4}$CSIRO Astronomy \& Space Science/NASA, Canberra Deep Space Communication Complex, PO Box 1035, Tuggeranong ACT 2901, Australia \\
$^{5}$University of Western Sydney, Locked Bag 1797, Penrith 2751 NSW, Australia \\
$^{6}$International Centre for Radio Astronomy Research, Curtin University, Perth, WA 6845, Australia \\
$^{7}$Sydney Institute for Astronomy, School of Physics, The University of Sydney, New South Wales 2006, Australia \\
$^{8}$School of Mathematics and Physics, University of Tasmania, Private Bag 37, Hobart 7001, Australia \\
$^{9}$Astrophysics Research Centre, School of Mathematics and Physics, Queen’s University Belfast, Belfast BT7 1NN, UK}

\begin{document}

\date{Accepted for publication in MNRAS, July 3 2014}

\pagerange{\pageref{firstpage}---\pageref{lastpage}} \pubyear{2014}

\maketitle

\label{firstpage}

\begin{abstract}
We present new detections of cyanodiacetylene (HC$_{5}$N) toward hot molecular cores, observed with the Tidbinbilla 34 m radio telescope (DSS--34). In a sample of 79 hot molecular cores, HC$_{5}$N was detected towards 35. These results are counter to the expectation that long chain cyanopolyynes, such as HC$_{5}$N, are not typically found in hot molecular cores, unlike their shorter chain counterpart HC$_{3}$N. However it is consistent with recent models which suggest HC$_{5}$N may exist for a limited period during the evolution of hot molecular cores.
\end{abstract}


\section{Introduction}
\noindent Molecular clouds support a rich organic chemistry which can be used to ascertain the progress of high-mass star formation within them. High mass star formation begins when dense (\textless10$^{4}$\  \rm{cm}$^{-3}$) clumps of cold gas and dust inside molecular clouds collapse under their own gravity and form cold  (\textless10\,K) cores. This stage proceeds for $\sim$5$\times$10$^{4}$ years. An increase in temperature, caused by the central protostar, and continuing collapse progresses the cloud to the dense ($\sim$10$^{6}$ \rm{cm}$^{-3}$), hot (\textgreater100\,K, typically $\sim$200\,K) core stage. Ices that were frozen on to dust grains in the previous stage are evaporated, resulting in a time-dependent chemistry in which gas phase reactions produce complex molecules \citep{b7}. This stage proceeds on a time-scale of $\sim$5$\times$10$^{4}$ to $\sim$1$\times$10$^{5}$ years. Methanol (CH$_{3}$OH) masers, exclusive markers of high mass star formation (\citealt{b16}, \citealt{b41}) also switch on in this stage \citep{b32}. \\ \let\thefootnote\relax\footnote{$^{\star}$E-mail: claire.elise.green@gmail.com}

Hot core chemistry is typically dominated by saturated molecules \citep{b17} and methyl cyanide (CH$_{3}$CN) is considered a key tracer of the hot core stage \citep{b7}. Cyanodiacetylene (HC$_{5}$N) is an unsaturated, complex organic molecule (a long chain cyanopolyyne, HC$_{2n+1}$N) associated with the early stages of star formation (e.g.  \citealt{b14}, \citealt{b21}, \citealt{b37}).  Although readily detected in cold (10K), dark (10$^{4}$ cm$^{-3}$) molecular clouds, the detection of HC$_{5}$N in warmer gas has proven difficult.  Recently, Sakai et al. (2008, 2009) have shown that a ‘warm carbon-chain chemistry’ (WCCC) occurs in warmer ($\sim$20 K), denser ($\sim$10$^{6}$ cm$^{-3}$) gas. In such gas it appears that the increase in grain temperature is sufficient to desorb the most lightly bound molecules, CO, N$_{2}$ and CH$_{4}$, from grain ices. The subsequent chemical processing of the evaporated CH$_{4}$, first to acetylene, C$_{2}$H$_{2}$, and its derivatives and then to diacteylene, C$_{4}$H$_{2}$, and larger carbon-chain molecules drives a complex organic chemistry \citep{b44}.  HC$_{5}$N has been detected in two WCCC sources, L1527 and IRAS15398--3359, with column densities a factor of 10--20 smaller than that in TMC--1. Its formation in these sources has not yet been explained as it was not discussed in the papers by Aikawa et al. (2008,  2012).  HC$_{5}$N is not typically associated with the more evolved hot core stage of high mass star formation unlike HC$_{3}$N \citep{b28}. However, rotational transitions of HC$_{5}$N have been observed within the Orion Molecular Cloud \citep{b23} and the Sagittarius B2 (Sgr B2) molecular cloud (\citealt{b3}; \citealt{b1}; \citealt {b23}). \\

HC$_{5}$N was not expected to be detected in hot cores due to the efficiency of hydrogenation on grain surfaces. It is very likely that both HC$_{5}$N, and its precursor molecules diacetylene (HC$_{4}$H) and its isomer butatrienylidene (H$_{2}$CCCC), will undergo addition reactions with hydrogen atoms in the ice mantles  that form in cold, dense gas prior to the onset of star formation. Earlier models of hot core chemistry found the formation of the precursors to be inefficient in the hot core and HC$_{5}$N abundances were predicted to be small \citep{b15}. \\

More recent chemical modelling, such as by \citet{b7}, using more complex chemical networks for the synthesis of large carbon chain molecules shows that the precursor molecules can, in fact, be formed efficiently in the hot core stage, driven by the high abundance of acetylene (C$_{2}$H$_{2}$) in the hot gas and predict that HC$_{5}$N can form and exist under the conditions of hot cores. The two major formation routes to HC$_{5}$N  involve H$_{2}$CCCC  \citep{b42} and HC$_{4}$H \citep{b9} and can be written in combined form as:
\begin{equation}
\label{eq:formation}
\rm{CN + C_{4}H_{2} \rightarrow HC_{5}N + H}
\end{equation}
Fukuzawa et al. (1998) showed that the reactions of CN with the higher polyacetylenes to form larger cyanopolyynes were also exothermic and have led to the suggestion that cyanohexatriyne (HC$_{7}$N)  and cyanooctatetrayne (HC$_{9}$N) may also be detectable in hot cores \citep{b7}.\\

\noindent To test these chemical models of star formation, there is a need to identify chemical species at different evolutionary stages, establishing a chemical clock (e.g. \citealt{b22}). For example,  \citet{b7} modelled an evolution of HC$_{5}$N to the longer chain cyanopolyynes, HC$_{7}$N and  HC$_{9}$N suggesting that detection of HC$_{5}$N indicates a relatively younger population of sources. \\

This paper presents new observations of HC$_{5}$N towards 79 hot molecular cores. The observations and data reduction are described in Section 2; results including spectra are presented in Section 3; and Section 4 discusses the nature of detections and implications for the chemical evolution of star formation. Conclusions are presented in Section 5. \\

\section{Observations}
\noindent Observations of 79 hot molecular cores were made with NASA Deep Space Station 34 (DSS--34), a 34 m diameter radio telescope at Tidbinbilla, located near Canberra, Australia. These observations were made in service observing mode across 26 sessions from 2006 July to 2008 May. Observations were spread over the 2006--2008 period due to the nature of the Tidbinbilla telescope availability, which depends on NASA scheduling priorities. Data from one observation block (2008 February 22) were excluded due to poor baselines attributed to bad weather conditions. The J=12$\rightarrow$11 transition of HC$_{5}$N was observed at its rest frequency of 31.951777 GHz towards these 79 sources. Position-switching was used with a typical integration time of $\sim$1 minute per position in an OFF-ON-ON-OFF pattern. Sources were observed with total integration times of on average $\sim$20 minutes (for the entire OFF-ON-ON-OFF pattern). The beamwidth was 0.95$'$ and the pointing accuracy was 2$''$. The correlator was configured to give single polarisation data with 2048 channels across a 64 MHz bandwidth centred at 31.951777 GHz. This provides a velocity channel width of 0.29 kms$^{-1}$. The initial data collected in 2006 was scaled differently than the 2007 and 2008 data due to different observing procedures and set ups. Therefore the 2006 data was used only to verify detections and not to classify a source as a `detection'. The 2006 data has been excluded from the following discussion and all analyses. \\

\subsection{Source Selection}
\noindent 74 of the 79 sources were associated with 6.7 GHz methanol masers and the remaining five were initially considered `maserless' cores, offset from methanol maser sites by $\sim$1--18 arcmins. Two of the five `maserless' cores (G05.89--0.39 and G05.90--0.43) have since been identified as hosting 6.7 GHz methanol masers \citep{b5} and the term `maserless' core will henceforth refer to the remaining three cores, G00.26+0.01, G14.99--0.70 and G15.03--0.71.  Sources were selected to correspond to those of \citet{b19} which represent a subset of the Walsh et al. (1997, 1998, 2003) methanol maser, radio and sub-millimetre surveys. The maserless cores were selected as potential pre-hot core candidates. Two sources, Sgr B2 and G00.26+0.01, were observed as reference sources for the presence of HC$_{5}$N across 12 and eight epochs respectively. The other 77 sources were observed between one and six times dependent on telescope availability.  \\

\subsection{Data Reduction}
\begin{figure}
\caption{\textbf{Distribution of rms noise.} `All non-detections'  refers to all spectra across multiple epochs with a non-detection of HC$_{5}$N.  `All detections' refers to all spectra across multiple epochs with a detection of HC$_{5}$N.}
\includegraphics[width=\columnwidth]{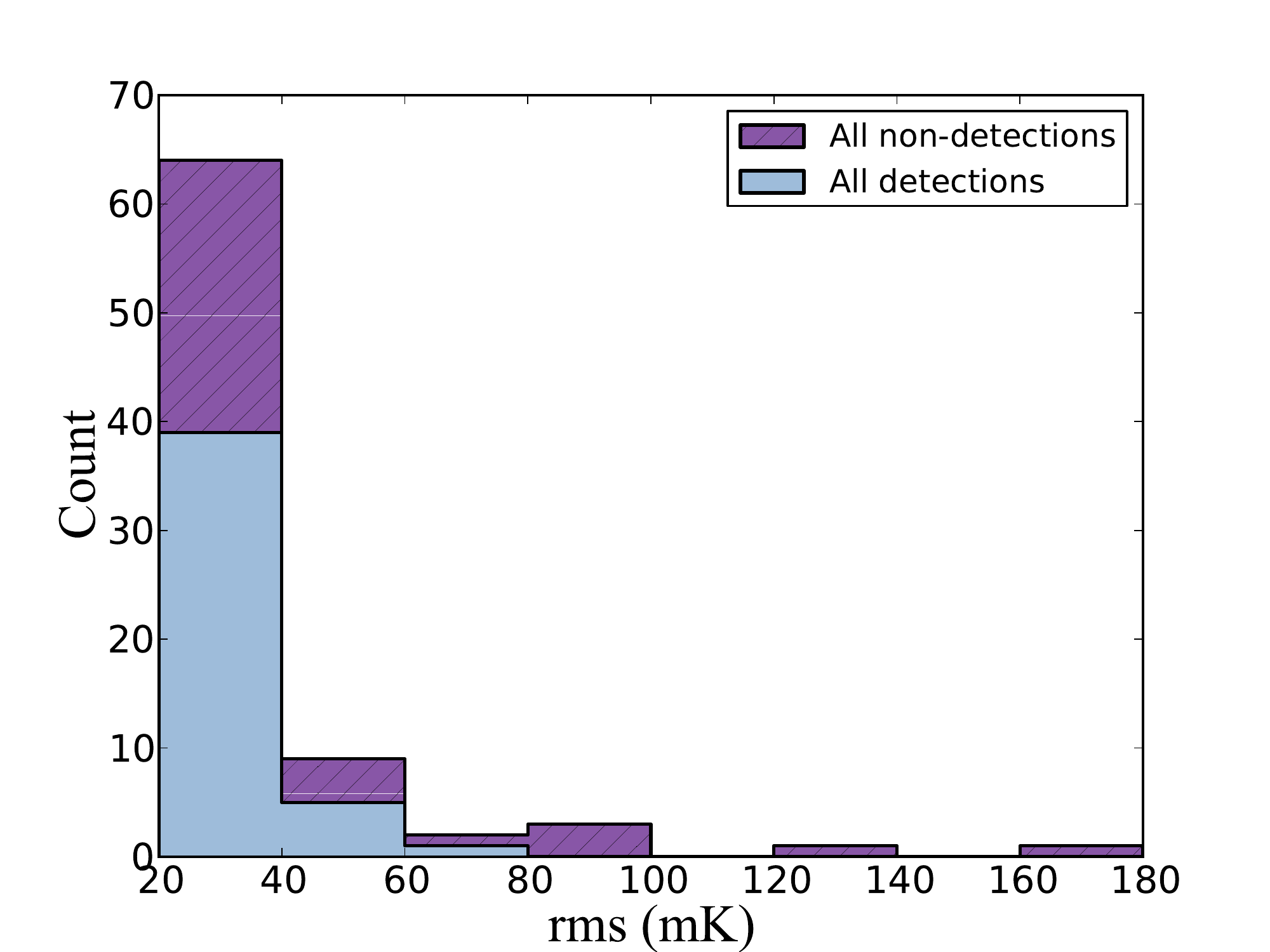}
\label{fig:noise_histogram}
\end{figure}

\noindent Data reduction was performed using the Australia Telescope National Facility (ATNF) Spectral Analysis Package (ASAP version 4.0.0)$^{1}$\let\thefootnote\relax\footnote{$^{1}$http://svn.atnf.csiro.au/trac/asap}. One spectrum was produced per source per epoch. Data was not combined across multiple epochs, as is standard, due to the unexpected scaling problems with the 2006 data. Although this was not an issue with the 2007 and 2008 data, the different epochs were reduced separately to ensure any other issues that did arise were isolated. For each spectrum, the reduction process involved taking the quotient of the source and reference scans and implementing a standard gain elevation correction. The antenna gain is empirically modelled by the following polynomial:
\begin{equation}
\rm{G(El) = R_{0} + R_{1}El + R_{2}(El^{2})} 
\end{equation} 
where \emph{El} is the elevation (in degrees), \emph{G(El)} gives the gain-elevation correction$^{2}$\let\thefootnote\relax\footnote{$^{2}$http://www.atnf.csiro.au/observers/docs/tid\_obs\_guide/tid\_obs\_guide\\
\_dss34.html}, \emph{R$_{0}$}=0.534289$\times$10$^{-1}$, \emph{R$_{1}$}=2.9831$\times$10$^{-3}$ and \emph{R$_{2}$}=--3.16376$\times$10$^{-5}$. The scans were aligned in the Local Standard of Rest Kinematic (LSRK) velocity frame and time weighted averaging was performed according to integration time. A baseline correction and standard Hanning smoothing across eight spectral channels were then applied. Final spectra were converted to the main beam temperature scale (T$_{\rm{MB}}$) by scaling by a factor of 0.7 to correct for beam efficiency (T. Kuiper \&  W. Veruttipong, private communication). Gaussian fitting using $\chi$$^{2}$ minimisation was then applied to the spectra to extract their peak width and centre velocity.  The nominal detection limit was three consecutive channels. The average rms for all spectra with detections was 21 mK.  Noise levels  in spectra of detections and non-detections were similar, with non-detections having an average rms value of 24 mK.  A  histogram comparing the noise levels of all spectra with detections across multiple epochs and all spectra with non-detections (which includes those for sources classified as having detections for different epochs) is presented in~\autoref{fig:noise_histogram}.  \\

We assume the HC$_{5}$N emission is optically thin based on the warm temperatures (100--200\,K) and likely small beam filling factor ($\textless$10 \%). \\

\section{Results}
\noindent Cyanodiacetylene (HC$_{5}$N) was detected in 35 of the 79 hot molecular cores to a 3$\sigma$ sensitivity limit of 54 mK. Detections were made at $\sim$3$\sigma$ to 17$\sigma$ where this signal to noise ratio was calculated by dividing the unrounded  peak T$_{\rm{MB}}$ by its associated unrounded error (calculated from the rms and Gaussian fit errors). The HC$_{5}$N detections included 33 of the maser associated cores and two out of the three `maserless' cores. The 35 detections include five weak detections at $\lesssim$3$\sigma$: G00.55--0.85, G008.14+0.23, G10.48+0.03, G11.50--1.49 and G25.71+0.04. Detections were made for  the vast majority of the 35 sources across multiple epochs. The epoch with the best signal to noise ratio and weather conditions was then selected. Detections are summarised in~\autoref{tab:detections}, non-detections are summarised in~\autoref{tab:non_detections} and HC$_{5}$N spectra are presented in~\autoref{fig:spectra}.  \\ 

\begin{figure*}
\caption{\textbf{Spectra of the 35 HC$_{5}$N detections.} Source name and epoch of observation are listed in the top left corner of each spectra. The dashed line represents the Gaussian fit, as per parameters listed in~\autoref{tab:detections}.}
\includegraphics[scale=0.9,clip=true, trim=0.8cm 0cm 0cm 1.5cm]{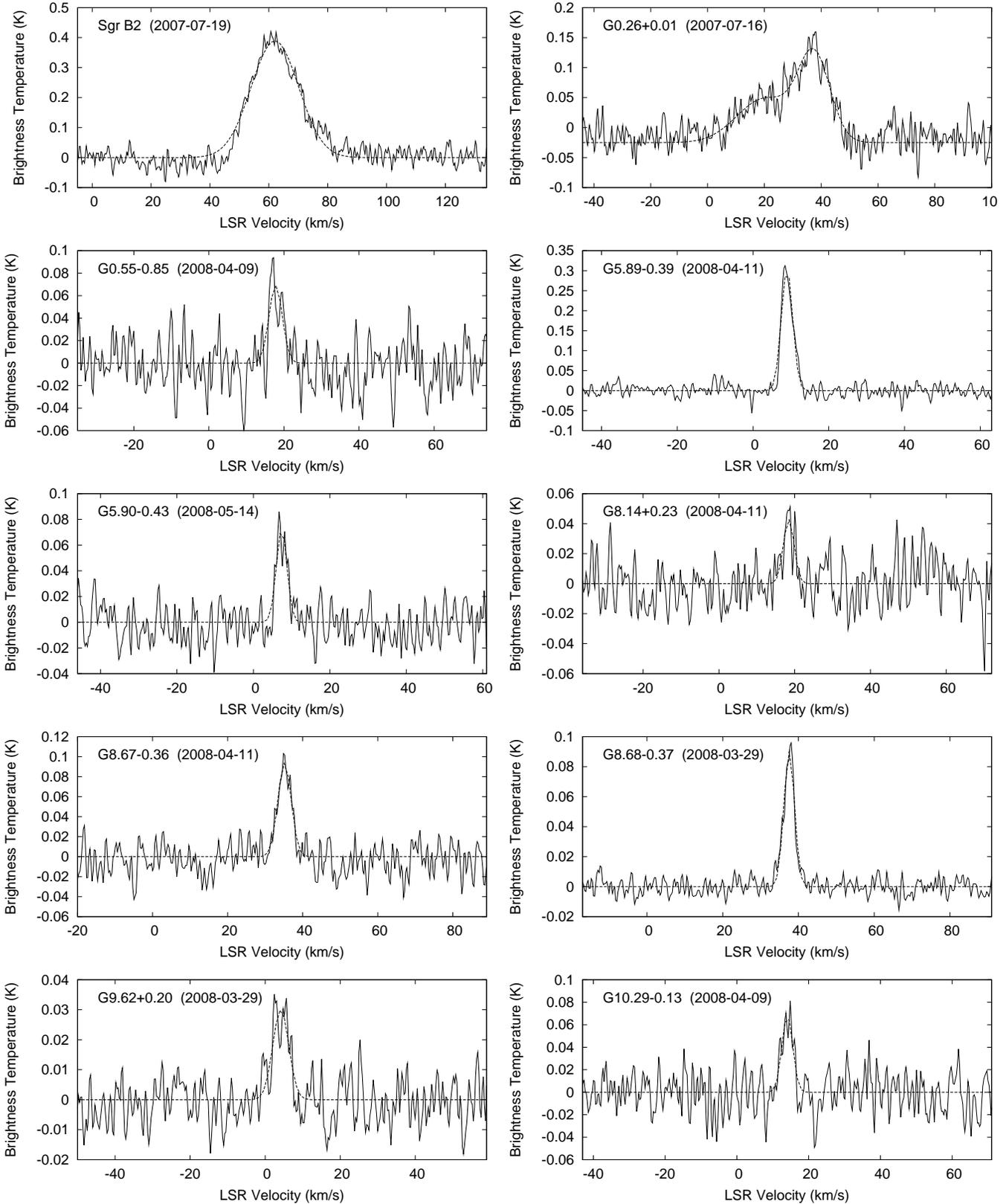}
\label{fig:spectra}
\end{figure*}

\begin{figure*}
\addtocounter{figure}{-1}
\caption{continued}
\includegraphics[scale=0.9,clip=true, trim=0.8cm 0cm 0cm 1.5cm]{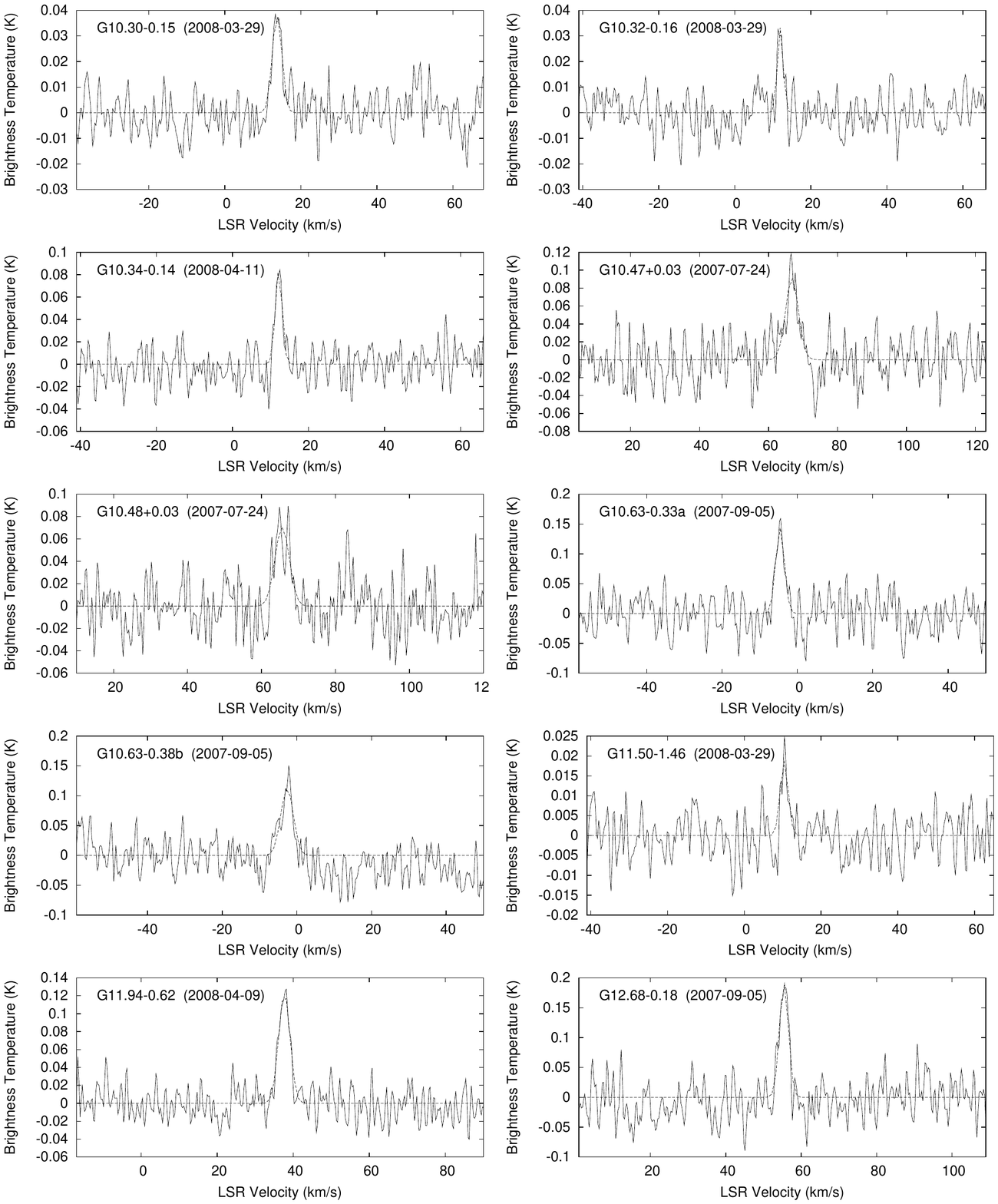}
\end{figure*}

\begin{figure*}
\addtocounter{figure}{-1}
\caption{continued}
\includegraphics[scale=0.9,clip=true, trim=0.8cm 0cm 0cm 1.5cm]{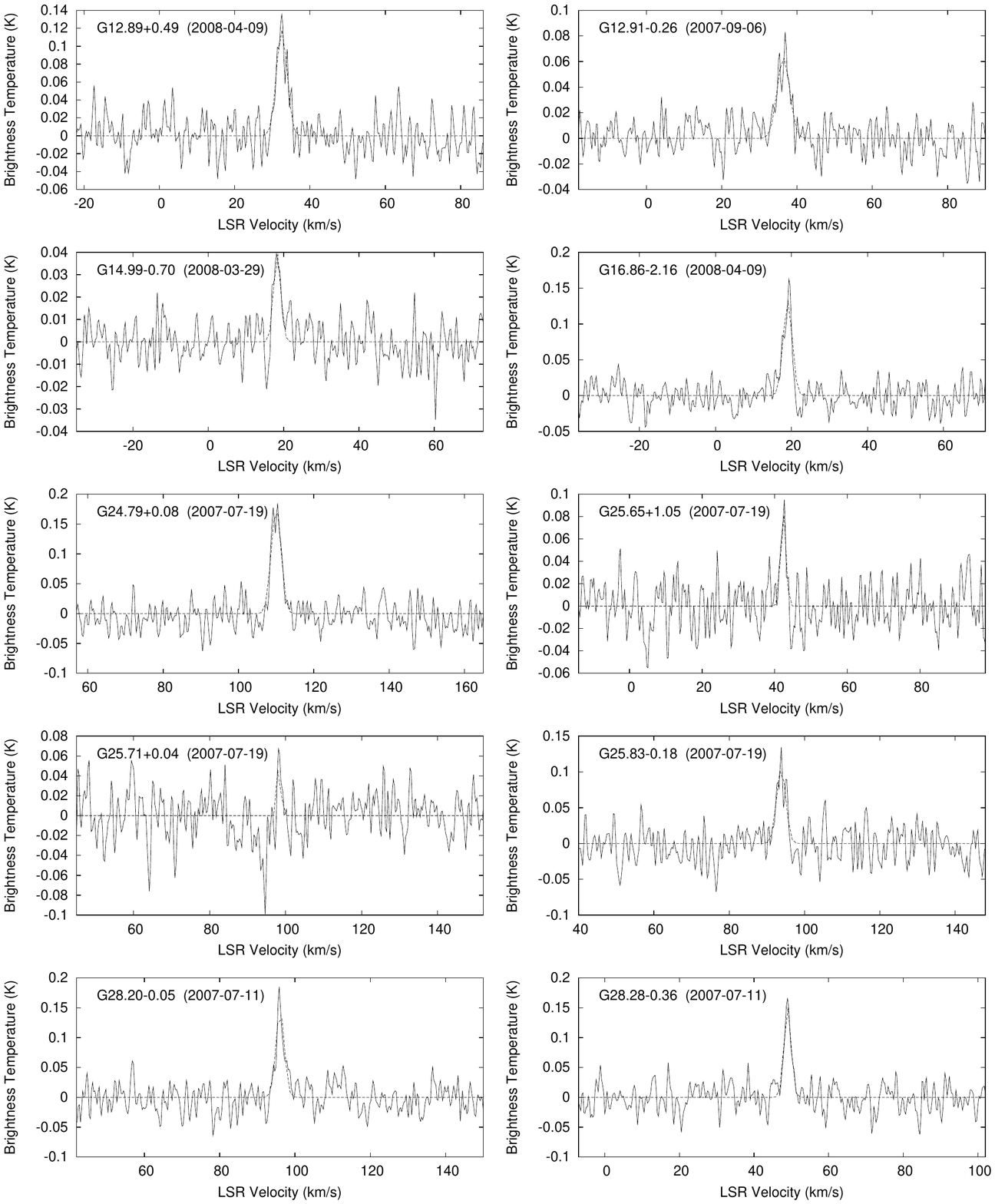}
\end{figure*}

\begin{figure*}
\addtocounter{figure}{-1}
\caption{continued}
\includegraphics[scale=0.9,clip=true, trim=0.8cm 0cm 0cm 1.5cm]{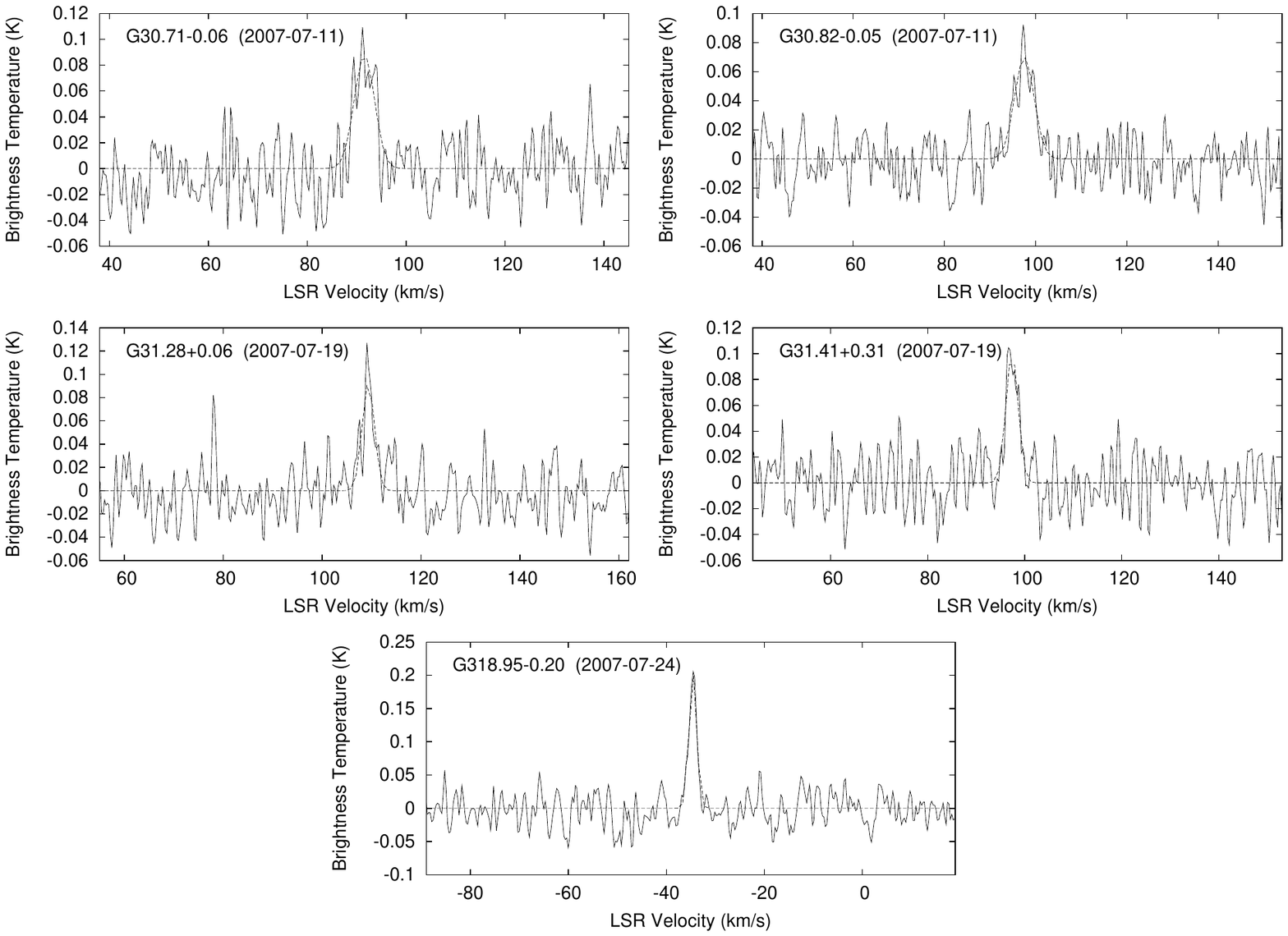}
\end{figure*}


\begin{figure}
\caption{\textbf{Distribution of peak brightness temperatures.} The red dashed line indicates the 3$\sigma$ sensitivity limit. Error bars represent the statistical (Poisson) error.}
\includegraphics[width=\columnwidth]{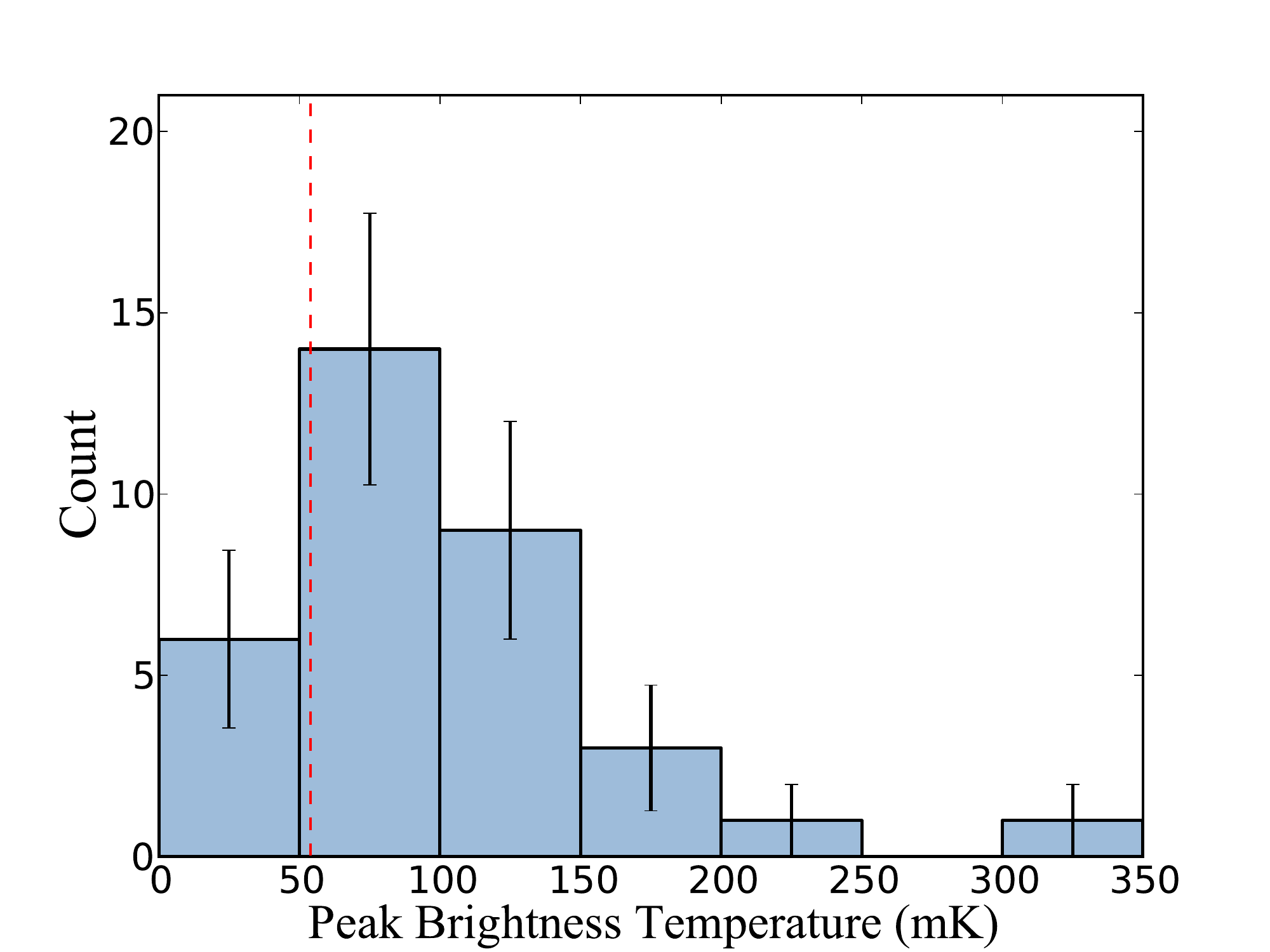}
\label{fig:brightness_temp_histogram}
\end{figure}

\begin{figure}
\caption{\textbf{Distribution of linewidths.} The red dashed line indicates the width of three channels. Error bars represent the statistical (Poisson) error. }
\includegraphics[width=\columnwidth]{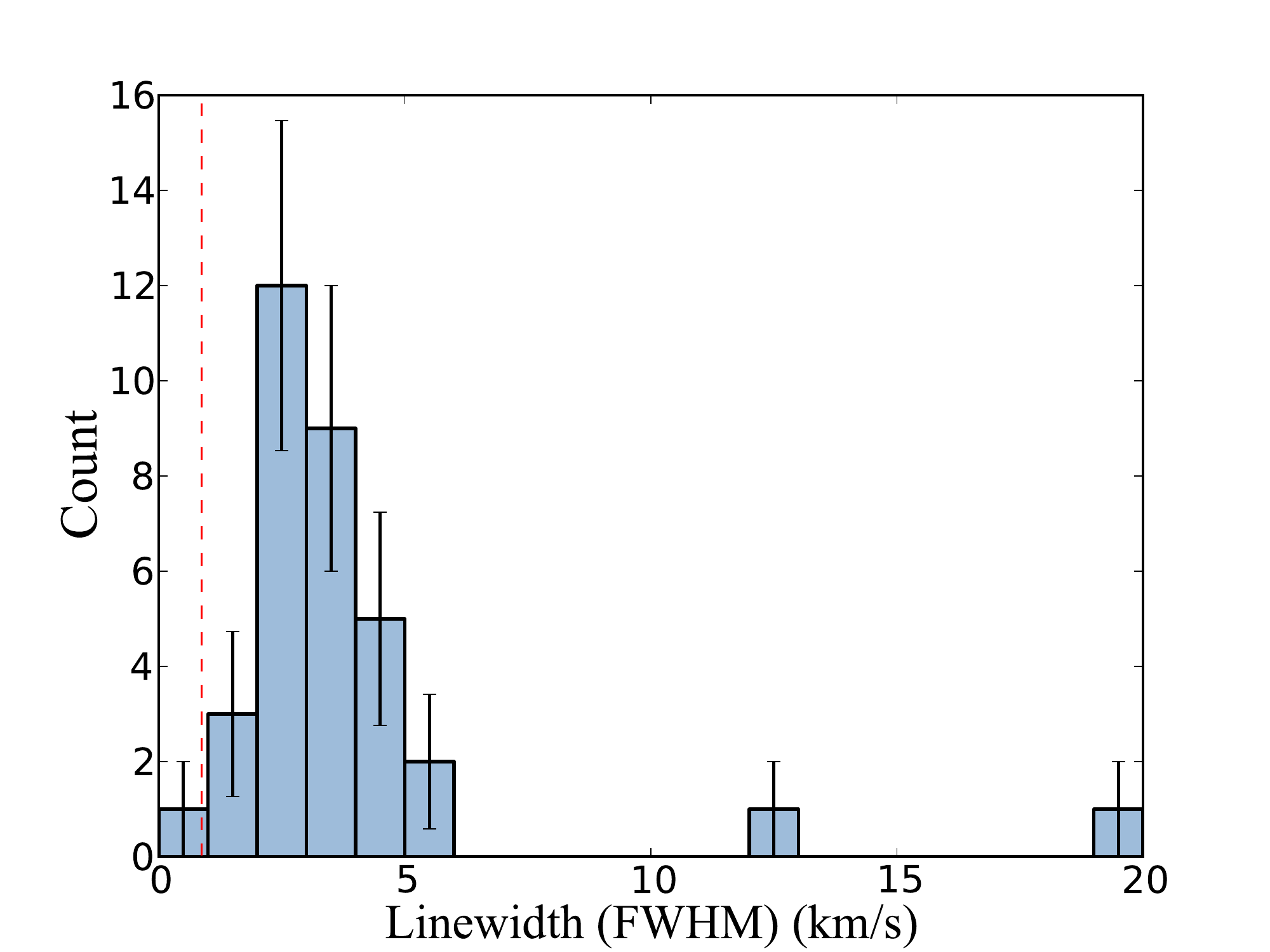}
\label{fig:FWHM_histogram}
\end{figure}

\begin{figure}
\caption{\textbf{HC$_{5}$N detections and kinematic distance.}  The count per bin has been normalised such that the total number of sources in each bin is one. Distance is not a factor in detection statistics. }
\includegraphics[width=\columnwidth]{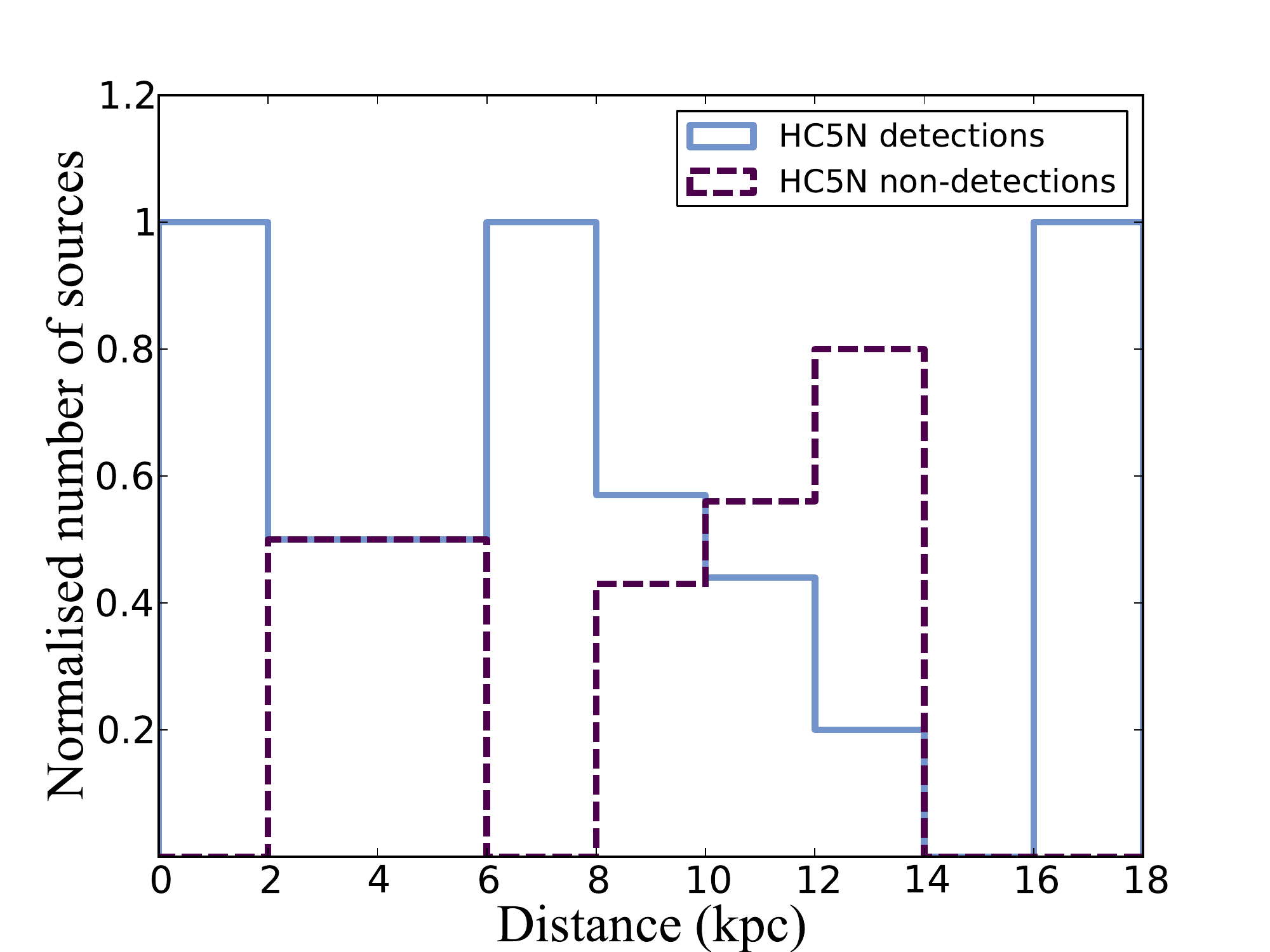}
\label{fig:distance_histogram}
\end{figure}

\begin{figure}
\caption{\textbf{HC$_{5}$N detections and methanol peak flux density.}  The count per bin has been normalised such that the total number of sources in each bin is one. Methanol peak flux density is not a factor in detection statistics.}
\includegraphics[width=\columnwidth]{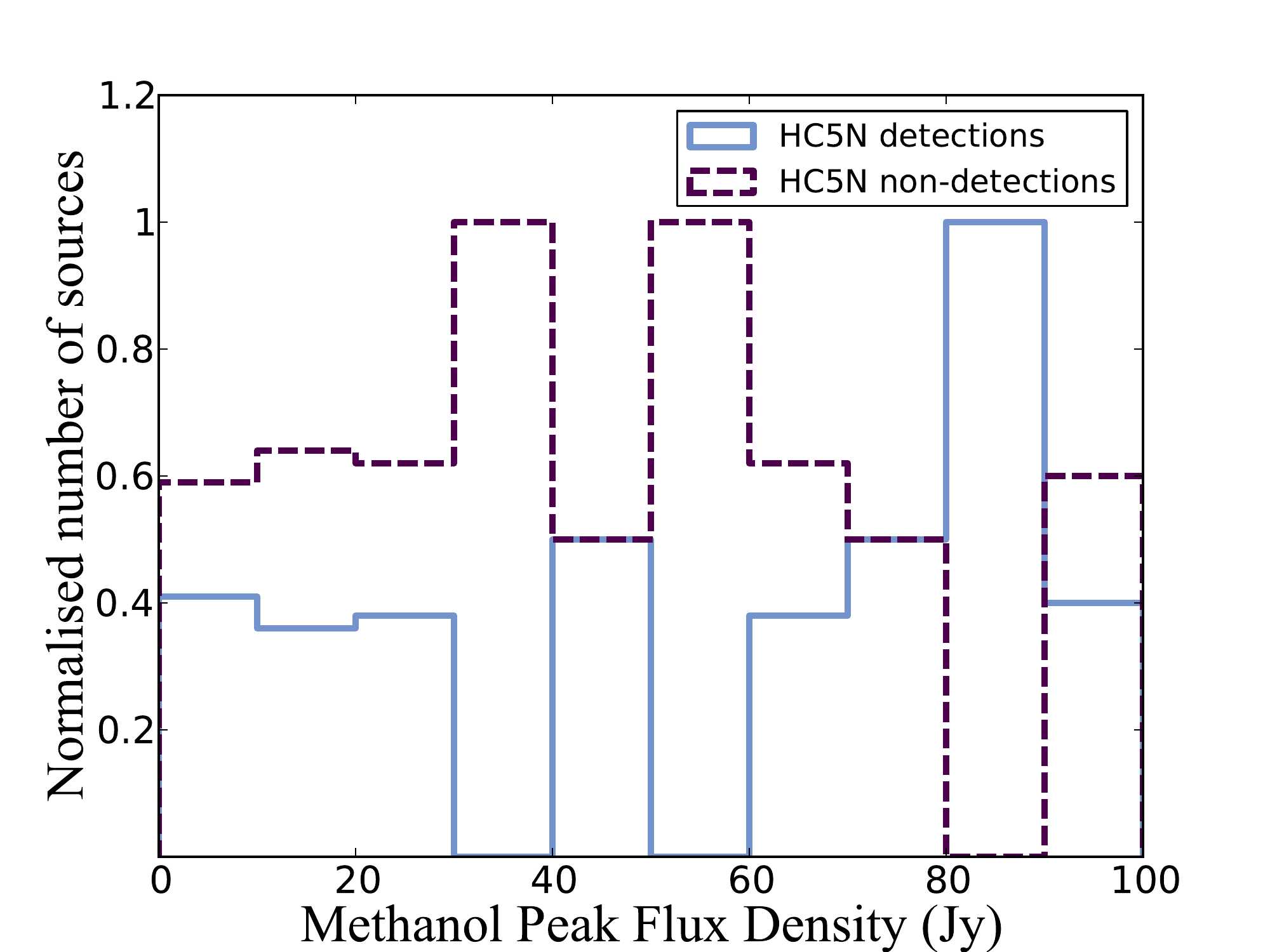}
\label{fig:methanol_histogram}
\end{figure}

The peak brightness temperature varied between 18 mK (taken to be the one $\sigma$ sensitivity limit, giving the 3$\sigma$ sensitivity limit of 54 mK) and 300 mK (excluding Sgr B2). The majority of sources had HC$_{5}$N peak brightness temperatures between 18 and 100 mK. The median peak brightness temperature was 91 mK. 
The brightest source, excluding Sgr B2, was G05.89--0.39 (300 mK). The distribution of peak brightness temperatures is shown in~\autoref{fig:brightness_temp_histogram}. \\

Linewidths (FWHM) varied between 1.0 and 5.1 kms$^{-1}$ (excluding Sgr B2 and G0.26+0.01, both broadened due to their location in the chaotic Galactic centre). The typical linewidth was 2--3 kms$^{-1}$ and the median was 3.3 kms$^{-1}$. The source with the widest linewidth, excluding Sgr B2 and G0.26+0.01,  was G30.82--0.05 which had a linewidth of 5.1 kms$^{-1}$. The distribution of the linewidths is shown in~\autoref{fig:FWHM_histogram}.

\begin{table*}
\setlength{\abovecaptionskip}{0pt}
\setlength{\belowcaptionskip}{0pt}
\caption{\textbf{The 35 HC$_{5}$N detections.} Column 1 is the Source Name; Column 2 is the Date of Observation (sources were observed across multiple epochs, the date of observation listed is that of the `best' individual detection determined by signal to noise ratio and weather conditions); Column 3 is the Peak Brightness Temperature (corrected on to the main beam scale with errors calculated from the rms and Gaussian fit error); Column 4 is the LSRK Velocity of the peak of the spectra; Column 5 is the Linewidth (FWHM); Column 6 is the Kinematic Distance; Column 7 is the Methanol Peak Flux Density; Column 8 indicates presence of OH masers with an `o' and water masers with a `w' (all sources are methanol masers except the three maserless cores); Column 9 indicates the molecules detected, with `c' indicating CH$_{3}$CN (\citealt{b19}, except for G11.50$-$1.49), `a' indicating HC$_{3}$N (reference for this molecule and for the CH$_{3}$CN molecule of source G11.50$-$1.49, is MALT90 Survey data available on the Australia Telescope Online Archive (ATOA)$^{3}$), `b' indicating HCN, `d' indicating HNC (Purcell et al. 2006, 2009. Additional reference is MALT90 data cubes). A $^{\dagger}$ indicates `maserless' cores \citep{b19}. References for quoted peak flux densities, distances and masers are:
$^{a}$\citet{b5},
$^{b}$\citet{b10},
$^{c}$\citet{b20},
$^{d}$\citet{b12},
$^{f}$\citet{b11},
$^{g}$\citet{b29},
$^{i}$\citet{b34},
$^{j}$\citet{b35},
$^{k}$\citet{b36},
$^{l}$\citet{b33}.
A $^{h}$ indicates distance is the kinematic distance calculated from the flat rotation curve with rotation velocity $\theta$=246 kms$^{-1}$ and solar distance R$_{\odot}$=8.4 kpc  (\citet{b29}; \citet{b30}; \citet{b31}).
}
\label{tab:detections}
\begin{tabular}{ | l| l |c |c | c| c| c|c|c|}
\hline
Source Name & Date & Peak T$_{\rm MB}$ & Vel$_{\rm LSRK}$  & Linewidth & Distance & S$_{\rm Peak \: Methanol}$ & Masers & Other  \\ 
& & & & (FWHM) & & & & Molecules\\ 
 & & (mK) & (km\,s$^{-1}$) & (km\,s$^{-1}$) &  (kpc) & (Jy) & & \\ 
\hline
Sgr B2 & 2007-07-19  & 390 $\pm$ 30 & 62.1   $\pm$  0.3 &  19.0 $\pm$ 0.4 & 7.8$^{g}$ & $-$ & o$^{j}$,w$^{j}$ & $-$\\ 
G00.26+0.01$^{\dagger}$ & 2007-07-16  & 140 $\pm$ 30 & 37.5  $\pm$  0.5 &  12.7 $\pm$ 0.9 & 8.4$^{h}$ & $-$ & & c,a,b,d   \\ 
 &   & 67 $\pm$ 20 & 20.8 $\pm$ 1.9 &  22.4 $\pm$ 4.3  & & & \\ 
G00.55$-$0.85 & 2008-04-09  & 69 $\pm$ 20 & 17.8   $\pm$  0.4 &  4.2 $\pm$ 0.8 & 8.4$^{h}$ & 61.9$^{a}$ & o$^{j}$,w$^{j}$ & c \\ 
G05.89$-$0.39 & 2008-04-11  & 300 $\pm$ 20 & 9.0   $\pm$  0.3 &  3.5 $\pm$ 0.3 & 1.9$^{f}$ & 0.5$^{a}$ & o$^{j}$,w$^{j}$ & c,a,b,d \\ 
G05.90$-$0.43 & 2008-05-14  & 72 $\pm$ 10 & 7.4   $\pm$  0.3 &  3.3 $\pm$ 0.4 & 1.6$^{f}$ & 6.2$^{a}$& w$^{j}$ & c,a,b,d \\ 
G08.14+0.23 & 2008-04-11  & 42 $\pm$ 20 & 18.4   $\pm$  0.4 &  3.4 $\pm$ 0.6 & 3.2$^{f}$ & 11.4$^{b}$ & w$^{j}$ & c,b,d \\ 
G08.67$-$0.36 & 2008-04-11  & 93 $\pm$ 10 & 35.2   $\pm$  0.3 &  4.2 $\pm$ 0.3 & 4.4$^{f}$ & 10.0$^{b}$ & o$^{j}$,w$^{j}$ & c,a,b,d \\ 
G08.68$-$0.37 & 2008-03-29  & 90 $\pm$ 06 & 37.6   $\pm$  0.3 &  3.3 $\pm$ 0.3 & 4.5$^{f}$ & 102.0$^{b}$ & w$^{j}$ & c,a,b,d \\ 
G09.62+0.20 & 2008-03-29  & 30 $\pm$ 07 & 4.2   $\pm$  0.4 &  5.0 $\pm$ 0.6 & 5.2$^{f}$ & 5239.9$^{b}$ & o$^{j}$,w$^{j}$ & c,a,b,d  \\ 
G10.29$-$0.13 & 2008-04-09  & 68 $\pm$ 20 & 14.1   $\pm$  0.3 &  3.6 $\pm$ 0.5 & 2.1$^{h}$ & 7.2$^{b}$& w$^{j}$ & c,a,b,d  \\ 
G10.30$-$0.15 & 2008-03-29  & 38 $\pm$ 08 & 13.7   $\pm$  0.3 &  3.0 $\pm$ 0.4 & 2.1$^{h}$ & 0.9$^{b}$ & & c,a,b,d \\ 
G10.32$-$0.16 & 2008-03-29  & 33 $\pm$ 08 & 12.0   $\pm$  0.3 &  1.8 $\pm$ 0.4 & 1.9$^{h}$ & 90.1$^{b}$ & w$^{j}$ &  c,a,b,d\\ 
G10.34$-$0.14 & 2008-04-11  & 84 $\pm$ 20 & 12.2   $\pm$  0.3 &  2.1 $\pm$ 0.4 & 1.9$^{h}$ & 15.1$^{b}$ & w$^{j}$ & c,a,b,d\\ 
G10.47+0.03 & 2007-07-24  & 91 $\pm$ 20 & 66.8   $\pm$  0.4 &  4.4 $\pm$ 0.5 & 11.2$^{f}$ & 28.0$^{b}$ & o$^{j}$,w$^{j}$ & c,a,b,d \\ 
G10.48+0.03 & 2007-07-24  & 71 $\pm$ 20 & 65.6   $\pm$  0.4 &  4.6 $\pm$ 0.7 & 11.4$^{f}$ & 22.5$^{b}$ & o$^{j}$,w$^{j}$ & c,a,b,d \\ 
G10.63$-$0.33a & 2007-09-05  & 140 $\pm$ 30 & --4.6   $\pm$  0.3 &  2.7 $\pm$ 0.4 & 5.2$^{f}$ & 5.0$^{b}$ & w$^{j}$ & a,b,d \\ 
G10.63$-$0.38b & 2007-09-05  & 110 $\pm$ 30 & --2.6   $\pm$  0.3 &  4.0 $\pm$ 0.6 & 17.0$^{h}$ & 4.2$^{b}$ & & c,a,b,d \\ 
G11.50$-$1.49 & 2008-03-29  & 18 $\pm$ 06 & 10.4   $\pm$  0.3 &  2.5 $\pm$ 0.5 & 1.6$^{f}$ & 68.4$^{b}$ & w$^{j}$ &  c,a,b,d\\ 
G11.94$-$0.62 & 2008-04-09  & 120 $\pm$ 20 & 37.6   $\pm$  0.3 &  3.5 $\pm$ 0.3 & 3.7$^{f}$ & 42.9$^{b}$ & &  c,a,b,d\\ 
G12.68$-$0.18 & 2007-09-05  & 190 $\pm$ 30 & 55.4   $\pm$  0.3 &  2.8 $\pm$ 0.4 & 4.5$^{h}$ & 351.0$^{b}$ & o$^{j}$,w$^{j}$ &  c,a,b,d\\ 
G12.89+0.49 & 2008-04-09  & 120 $\pm$ 20 & 32.6   $\pm$  0.3 &  3.3 $\pm$ 0.4 & 2.3$^{f}$ & 68.9$^{b}$ & o$^{j}$,w$^{j}$ & c,a,b,d \\ 
G12.91$-$0.26 & 2007-09-06  & 63 $\pm$ 20 & 36.4   $\pm$  0.3 &  4.0 $\pm$ 0.2 & 3.5$^{h}$ & 269.1$^{b}$ & o$^{j}$,w$^{j}$ &  c,a,b,d\\ 
G14.99$-$0.70$^{\dagger}$ & 2008-03-29  & 39 $\pm$ 09 & 18.3   $\pm$  0.3 &  2.3 $\pm$ 0.5 & 2.0$^{h}$ & $-$ & & c\\ 
G16.86$-$2.16 & 2008-04-09  & 130 $\pm$ 20 & 19.0   $\pm$  0.3 &  2.8 $\pm$ 0.4 & 1.9$^{h}$ & 28.9$^{b}$ & & c \\ 
G24.79+0.08 & 2007-07-19  & 180 $\pm$ 30 & 110.0   $\pm$  0.3 &  3.0 $\pm$ 0.3 & 9.6$^{f}$ & 97.0$^{c}$ & o$^{k}$,w$^{k}$ & c \\ 
G25.65+1.05 & 2007-07-19  & 84 $\pm$ 20 & 42.4   $\pm$  0.3 &  1.7 $\pm$ 0.4 & 12.5$^{f}$ & 178.0$^{c}$ & & c \\ 
G25.71+0.04 & 2007-07-19  & 69 $\pm$ 30 & 98.3   $\pm$  0.3 &  1.0 $\pm$ 0.4 & 10.1$^{f}$ & 364.0$^{c}$ & & c \\ 
G25.83$-$0.18 & 2007-07-19  & 110 $\pm$ 20 & 93.8   $\pm$  0.3 &  2.9 $\pm$ 0.4 & 5.0$^{f}$ & 70.0$^{c}$ & & c \\ 
G28.20$-$0.05 & 2007-07-11  & 140 $\pm$ 20 & 96.1   $\pm$  0.3 &  2.6 $\pm$ 0.4 & 9.8$^{f}$ & 3.3$^{c}$ & o$^{i}$ & c\\ 
G28.28$-$0.36 & 2007-07-11  & 150 $\pm$ 20 & 49.0   $\pm$  0.3 &  2.2 $\pm$ 0.4 & 3.0$^{h}$ & 62.0$^{c}$ & & c \\ 
G30.71$-$0.06 & 2007-07-11  & 88 $\pm$ 20 & 91.5   $\pm$  0.4 &  4.9 $\pm$ 0.8 & 4.9$^{h}$ & 87.0$^{c}$ & & c \\ 
G30.82$-$0.05 & 2007-07-11  & 70 $\pm$ 20 & 97.5   $\pm$  0.3 &  5.1 $\pm$ 0.5 & 4.9$^{f}$ & 18.0$^{c}$ & & c \\ 
G31.28+0.06 & 2007-07-19  & 92 $\pm$ 20 & 109.3   $\pm$  0.3 &  2.9 $\pm$ 0.5 & 5.8$^{f}$ & 71.0$^{c}$ & & c \\ 
G31.41+0.31 & 2007-07-19  & 100 $\pm$ 20 & 97.4   $\pm$  0.3 &  3.0 $\pm$ 0.4 & 6.6$^{f}$ & 11.0$^{c}$ & & c \\ 
G318.95$-$0.20 & 2007-07-24  & 200 $\pm$ 20 & --34.5   $\pm$  0.3 &  1.9 $\pm$ 0.3 & 10.6$^{f}$ & 569.2$^{d}$ & o$^{l}$,w$^{j}$ & c,a,b,d\\ 
\hline
\end{tabular}
\end{table*}
\newpage

\begin{table*}
\setlength{\abovecaptionskip}{0pt}
\setlength{\belowcaptionskip}{0pt}
\begin{center}
\caption{\textbf{Non-detections of HC$_{5}$N.} These sources fall below the 3$\sigma$ detection limit of 54 mK. Column 1 is the Source Name; Column 2 is the Date of Observation; Column 3 is the Kinematic Distance; Column 4 is the Methanol Peak Flux Density; Column 5 indicates presence of OH masers with an `o' and water masers with a `w' (all sources are methanol masers except the three maserless cores); Column 6 indicates the molecules detected, with `c' indicating CH$_{3}$CN (\citealt{b19}), `a' indicating HC$_{3}$N (MALT90 Survey data), `b' indicating HCN, `d' indicating HNC (Purcell et al. 2006, 2009. Additional reference is MALT90 data cubes). A $^{\dagger}$ indicates `maserless' cores \citep{b19}. References for quoted peak flux densities, distances and masers are:
$^{a}$\citet{b5},
$^{b}$\citet{b10},
$^{c}$Pestalozzi et al. (2005),
$^{d}$\citet{b11},
$^{e}$\citet{b6},
$^{f}$\citet{b12},
$^{i}$\citet{b34},
$^{j}$\citet{b35},
$^{k}$\citet{b36}. 
}
\label{tab:non_detections}
\begin{tabular}{ | l| l |c |c |c| c| }
\hline
Source Name & Observation Dates & Distance & S$_{\rm Peak Methanol}$ & Masers & Other \\ 
 & & (kpc) & (Jy) & & Molecules\\ 
\hline
G00.21$-$0.00 & 2006-08-05, 2008-03-29, 2008-04-09, 2008-04-11  & $-$ & 3.3$^{a}$ & w$^{j}$ & \\ 
G00.32$-$0.20 & 2007-07-24  & $-$ & 62.60$^{a}$ & w$^{j}$ & c \\ 
G00.50+0.19 & 2006-07-27, 2007-10-01, 2007-12-27, 2008-05-14  & $-$ & 24.5$^{a}$ & o$^{j}$,w$^{j}$ & d \\ 
G00.84+0.19 & 2006-07-27, 2007-10-01, 2007-12-27  & $-$ & 6.6$^{a}$ & & c,d \\ 
G01.15$-$0.12 & 2006-07-27  & $-$ & 3.0$^{a}$ & & c \\ 
G02.54+0.20 & 2006-08-05, 2007-10-01, 2007-12-28  & $-$ & 29.4$^{a}$ & w$^{j}$ & \\ 
G06.54$-$0.11 & 2006-07-10, 2008-03-29,  2008-04-11  & 13.9$^{d}$ & 0.6$^{b}$ & w$^{j}$ & \\ 
G06.61$-$0.08 & 2007-05-29, 2007-07-24, 2008-05-14  & $-$ & 23.4$^{b}$ & w$^{j}$ & \\ 
G09.99$-$0.03 & 2006-08-05, 2007-05-29, 2007-07-19, 2008-04-11  & 12.0$^{d}$ & 67.6$^{b}$ & w$^{j}$ & c,b,d \\ 
G10.44$-$0.02 & 2007-06-15, 2007-07-24  & 11.0$^{d}$ & 24.3$^{b}$ & o$^{i}$,w$^{j}$ & c \\ 
G11.99$-$0.27 & 2006-07-27, 2008-03-29, 2008-04-11  & 11.7$^{d}$ & 1.9$^{b}$ & & \\ 
G12.03$-$0.03 & 2006-07-27, 2008-03-29, 2008-04-11  & 11.1$^{d}$ & 96.3$^{b}$ & & \\ 
G12.18$-$0.12 & 2007-06-15, 2007-09-05  & $-$ & 1.9$^{b}$ & & \\ 
G12.21$-$0.09 & 2007-06-15, 2007-09-05  & $-$ & 11.5$^{b}$ & & \\ 
G14.60+0.02 & 2007-06-16, 2007-12-27  & 2.8$^{d}$ & 2.3$^{b}$ & & c,d \\ 
G15.03$-$0.67 & 2007-09-11, 2007-12-27  & 2.3$^{d}$ & 47.5$^{b}$ & & c \\ 
G15.03$-$0.71$^{\dagger}$ & 2006-07-10, 2007-09-11, 2007-12-27  & $-$ & $-$ & & \\  
G19.36$-$0.03 & 2007-06-16, 2007-09-11, 2007-09-18  & 2.3$^{d}$ & 33.8$^{b}$ & & c \\ 
G19.47+0.17 & 2007-06-16, 2007-09-11, 2007-12-28  & $-$ & $-$ & & c \\ 
G19.49+0.15 & 2007-06-16, 2007-09-11, 2007-12-28, 2008-02-22, 2008-03-23, 2008-04-09  & 2.0$^{d}$ & 6.0$^{b}$ & & \\ 
G19.61$-$0.13 & 2007-06-16, 2007-09-11, 2007-12-28  & 12.1$^{d}$ & 12.5$^{b}$ & o$^{k}$ & c \\ 
G19.70$-$0.27 & 2007-06-16, 2007-12-28  & 12.6$^{d}$ & 10.0$^{b}$ & & \\ 
G21.88+0.01 & 2007-12-28, 2007-07-05, 2007-07-16  & $-$ & 15.0$^{c}$ & & \\ 
G22.36+0.07 & 2007-07-05, 2007-07-16, 2007-10-01  & 4.6$^{d}$ & 12.0$^{c}$ & & c\\ 
G23.26$-$0.24 & 2007-07-05, 2007-07-11, 2007-07-16  & $-$ & 4.4$^{c}$ & & \\ 
G23.44$-$0.18 & 2007-07-05, 2007-07-16  & 5.9$^{d}$ & 77.0$^{c}$ & o$^{k}$,w$^{k}$ & c \\ 
G23.71$-$0.20 & 2007-07-05, 2007-07-16, 2007-07-19  & 11.0$^{d}$ & 9.2$^{c}$ & & \\ 
G28.15$-$0.00 & 2007-07-05, 2007-07-19, 2007-10-01, 2007-12-28  & 5.3$^{d}$ & 34.0$^{c}$ & o$^{i}$ & c \\ 
G28.31$-$0.39 & 2007-07-11, 2007-10-01, 2007-12-28  & 10.4$^{d}$ & 62.0$^{c}$ & & \\ 
G28.83$-$0.25 & 2007-07-11  & 4.6$^{d}$ & 73.0$^{c}$ & &\\ 
G28.85$-$0.23 & 2007-07-11  & 5.4$^{d}$ & 1.9$^{c}$ & & \\ 
G29.87$-$0.05 & 2007-07-11  & $-$ & 67.0$^{c}$ & &\\ 
G29.96$-$0.02 & 2007-07-11  & 9.3$^{d}$ & 206.0$^{c}$ & & c \\ 
G29.98$-$0.04 & 2007-07-11, 2007-12-28, 2008-05-14  & 9.2$^{d}$ & 14.0$^{c}$ & & c\\ 
G30.59$-$0.04 & 2007-07-11, 2008-03-17, 2008-03-23, 2008-04-09, 2008-05-14  & 2.7$^{d}$ & 7.5$^{c}$ & & \\ 
G30.76$-$0.05 & 2007-07-11  & 4.8$^{d}$ & 68.0$^{c}$ & & c\\ 
G30.78+0.23 & 2007-07-19  & $-$ & 19.0$^{c}$ & & \\ 
G30.79+0.21 & 2007-07-19  & 9.9$^{d}$ & 23.0$^{c}$ & & c \\ 
G30.82+0.28 & 2007-07-11, 2007-07-19  & 5.6$^{d}$ & 8.0$^{c}$ & & c \\ 
G30.90+0.16 & 2007-07-11  & 5.6$^{d}$ & 95.2$^{c}$ & & c \\  
G316.81$-$0.06 & 2007-05-29, 2007-06-15, 2007-06-16, 2007-07-24, 2008-03-23, 2008-04-09  & 2.6$^{d}$ & 52.0$^{f}$ & o$^{j}$,w$^{j}$ &  c,a,b,d\\ 
G323.74$-$0.26 & 2007-05-29, 2007-06-15, 2007-07-24, 2008-03-23, 2008-04-09  & 2.8 $^{d}$& 3114.4$^{f}$ & o$^{j}$,w$^{j}$ & c,a,b,d \\ 
G331.28$-$0.190 & 2007-05-29, 2007-06-15,  2007-07-24,  2008-03-23, 2008-03-29, 2008-04-09  & 4.4$^{d}$ & 90.0$^{e}$ & o$^{j}$,w$^{j}$ & c,a,b,d\\ 
G332.73$-$0.62 & 2007-05-29, 2007-06-15, 2007-07-24, 2008-03-23, 2008-03-29, 2008-04-09  & 3.0$^{d}$ & 5.1$^{e}$& o$^{j}$,w$^{j}$ & c,b,d \\ 
\hline
\end{tabular}
\end{center}
\end{table*}
\newpage
\newpage
\vspace*{-5mm}
\section{Discussion} 
\subsection{Comparison with methanol maser attributes}
\noindent Properties of the HC$_{5}$N detections and non-detections were compared with the properties of the methanol masers with which they are associated.~\autoref{fig:distance_histogram} compares the distance to the methanol masers  with the normalised number of detections and non-detections.  As there are unequal numbers of detections and non-detections, normalisation sets the total number of detections and non-detections in each  bin to unity for ease of comparison.  We detect HC$_{5}$N across a range of distances and do not see evidence for a fall-off with distance. The majority (6 of the 8 distance bins) have no difference between detections and non-detections with (heliocentric) kinematic distance. This implies either the sample is too small (with results dominated by the statistical errors) or that the sensitivity of these observations is sufficient to detect sources to a distance of $\sim$18 kpc.~\autoref{fig:methanol_histogram} compares the peak flux density of the methanol masers  with the normalised number of detections and non-detections. Source Sgr B2 has been excluded from this figure as it represents a reference source for the presence of HC$_{5}$N and was used as a check of the observations. As one of the largest molecular clouds in the Galaxy, it does not offer good comparison to the rest of the sample and has thus been excluded from all such analyses. Five other sources G00.26+0.01, G00.55--0.85, G14.99--0.70, G15.03--0.71 and G19.47+0.17 were also excluded from this figure due to lack of available data. Again no clear relation was found in the sample, implying methanol peak flux density is not correlated with the absence or presence of HC$_{5}$N.   \footnotetext{$^{3}$atoa.atnf.csiro.au}\\

\subsection{Comparison with other molecules}
\noindent The sources with both HC$_{5}$N detections and non-detections were compared with the presence of other molecules. If HC$_{5}$N was detected, CH$_{3}$CN (5--4 and/or 6--5) (Purcell et al. 2006, 2009) was also detected in all but one source (G11.50--1.49). Both molecules are considered tracers of dense molecular clouds, but although the excitation conditions of these molecules are similar in density, they differ in temperature.  HC$_{5}$N and CH$_{3}$CN have similar critical densities$^{4}$ of $\sim$10$^{5}$ and $\sim$10$^{6}$  cm$^{-3}$ respectively, but the longer carbon chain HC$_{5}$N was expected to be destroyed at the higher temperatures that CH$_{3}$CN traces.\footnotetext{$^{4}$Critical densities have been calculated assuming a typical hot core temperature of 200\,K, a collisional partner of H$_{2}$ (giving an average velocity (v) of $\sim$1 kms$^{-1}$), and a typical collisional cross-section ($\sigma$) of 10$^{-16}$ cm$^{2}$. The reference for the Einstein A coefficients used for this calculation is the CDMS database (http://www.astro.uni-koeln.de/cgi-bin/cdmssearch). } As discussed in Section 1, CH$_{3}$CN is a tracer of the hot core stage of high mass star formation while HC$_{5}$N is more usually associated with the colder and earlier stages of star formation, so the detection of HC$_{5}$N in 35 hot molecular cores represents a departure from this paradigm. The detection of HC$_{5}$N in these hot molecular cores supports the results of the chemical modelling of \citet{b7}. They found that HC$_{5}$N could exist under the conditions of a hot molecular core, and suggest that if HC$_{5}$N exists in a region then HC$_{7}$N and HC$_{9}$N may also have the necessary conditions to form. They also suggest that the abundances of both of these molecules increase over time, although these species are not linked directly, as the rapid gas-phase chemistry begins after the destruction of the grain mantle species \citep{b7}. Thus the simultaneous detection of  HC$_{5}$N and CH$_{3}$CN is consistent with the \citet{b7} results.  However in 25 of the 44 HC$_{5}$N non-detection sources, CH$_{3}$CN was also detected, contrary to preliminary results presented in \citet{b7} which indicated that either both HC$_{5}$N and CH$_{3}$CN were detected together, or neither were detected.  Our analysis finds that this is not so and that CH$_{3}$CN is detected in both HC$_{5}$N detection and non-detection sources.  \\

It is possible that HC$_{5}$N can only exist for a limited period of time in these hot molecular cores. This may explain why HC$_{5}$N was detected in some of the hot molecular cores in the source list, but not all. The absence of HC$_{5}$N detections may also be due to any HC$_{5}$N transitions falling below the sensitivity limit. Alternatively the HC$_{5}$N may be emitting from another region outside the hot molecular core but still within the beam, for instance a cold, dense envelope surrounding the hot molecular core.  Further, higher sensitivity ATCA observations of HC$_{5}$N in these sources along with high sensitivity observations of a hot core tracer such as CH$_{3}$CN may be able to resolve this issue. Spatial comparison of the  HC$_{5}$N detections with archived MALT90 CH$_{3}$CN data was attempted to investigate whether a cold or warm envelope was the source of the HC$_{5}$N detections.  MALT90, however, is a mapping project and its detections were too weak to offer good spatial comparison for our HC$_{5}$N detections, which had significantly longer integration times. \\

\begin{figure*}
     \begin{center}
  \caption{\textbf{Molecular comparisons.} Peak brightness temperature of corresponding pairs for HC$_{5}$N, against N$_{2}$H$^{+}$, 1.2 mm flux, CH$_{3}$CN, Thermal CH$_{3}$OH, HCO$^{+}$ and HNC.	All peak brightness temperatures have been corrected on to the main beam scale. The crosses represent HC$_{5}$N detections, the circles represent HC$_{5}$N non-detections with main beam peak brightness temperature upper limits of 0.018 K, the one $\sigma$ detection limit. Reference for N$_{2}$H$^{+}$, CH$_{3}$CN, Thermal CH$_{3}$OH, HCO$^{+}$, HNC molecular data is Purcell et al. 2006, 2009.  The error margins (one $\sigma$) of each of these measurements are respectively $\sim$47 mK, $\sim$80 mK, $\sim$76 mK, $\sim$200 mK and $\sim$42 mK. Reference for 1.2 mm IR data is \citealt{b40}. The error margin (one $\sigma$) of these measurements is $\sim$150 mJy. Sources Sgr B2 and G28.85--0.23 have been excluded from this figure.}%
\label{fig:molec_comparisons}
	
\includegraphics[trim=1cm 0cm 0cm 3cm, clip=true,scale=0.8]{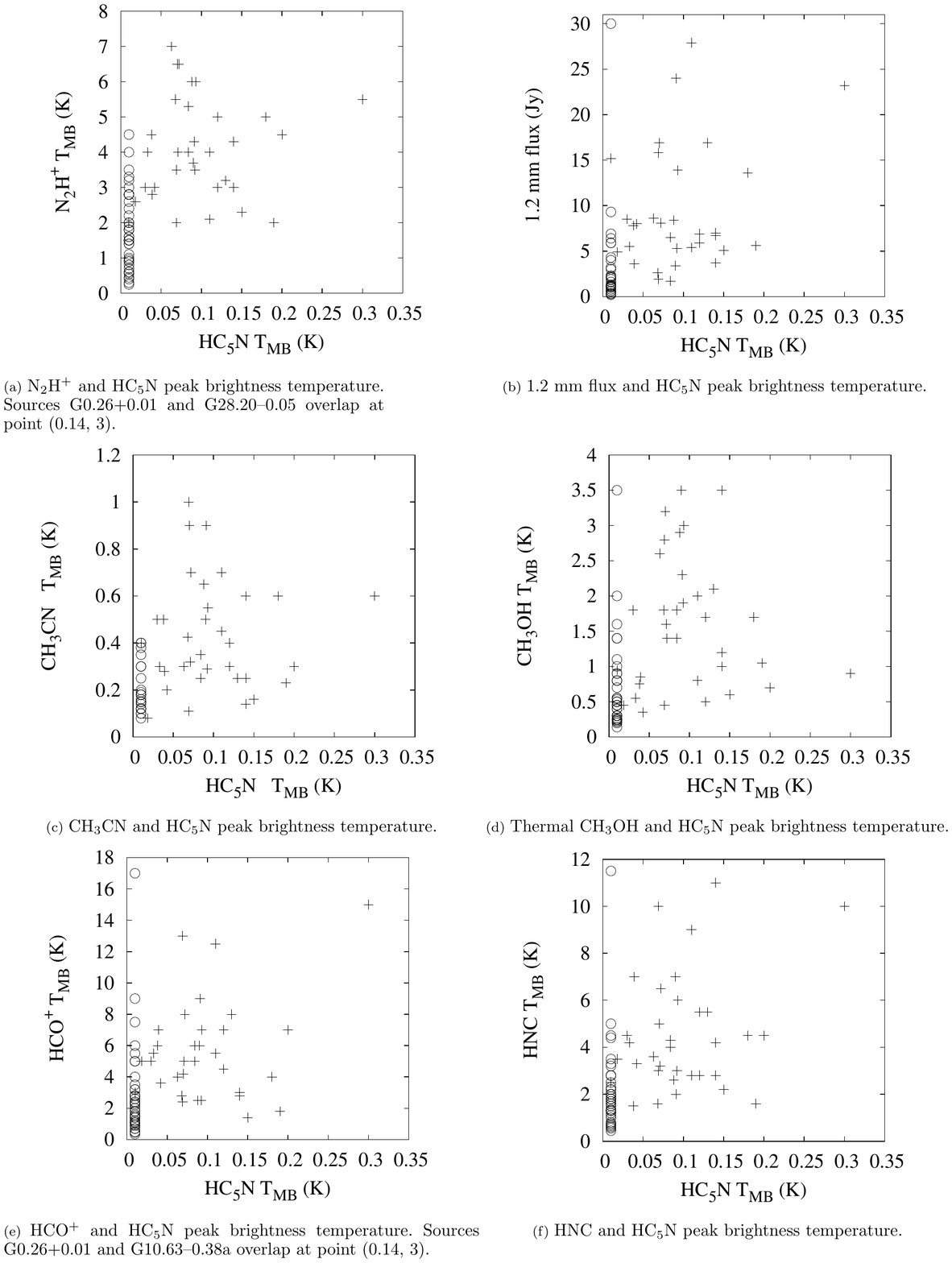}
	\end{center}
\end{figure*}

The peak brightness temperature of the molecular lines N$_{2}$H$^{+}$ (1--0), CH$_{3}$CN (5--4 and/or 6--5), thermal CH$_{3}$OH (2--1), HCO$^{+}$ (1--0), and HNC (1--0) (Purcell et al. 2006, 2009) are compared with the flux of HC$_{5}$N detections and non-detections in~\autoref{fig:molec_comparisons}. Source Sgr B2 has been excluded from this figure for reasons previously discussed along with source G28.85--0.23 which has been excluded due to lack of data for comparison. In addition sources G10.48+0.03 and G318.95--0.20 along with 11 of the non-detection sources have also been excluded from~\autoref{fig:molec_comparisons} b) for this reason. Approximately the same range of intensities for the detected molecule is spanned both when HC$_{5}$N has, and has not, been detected. However, in the case of N$_{2}$H$^{+}$, approximately half the detections of this molecule when HC$_{5}$N has not been seen fall well below their intensities when HC$_{5}$N has been detected. In other words,  N$_{2}$H$^{+}$ is significantly brighter in sources with HC$_{5}$N detections than without. HC$_{5}$N and N$_{2}$H$^{+}$ trace cold, dense gas with critical densities$^{4}$ of $\sim$10$^{5}$ and $\sim$10$^{6}$  cm$^{-3}$ respectively. N$_{2}$H$^{+}$, as well as tracing cold gas, can be found with significant optical depth in warm clouds \citep{b49}. Although N$_{2}$H$^{+}$ was a component of the reaction networks of \citet{b7}, they did not examine or comment on the behaviour of this ion or on relationships with HC$_{5}$N. Thus it cannot be fully determined whether this result is consistent with the modelling of \citet{b7}. Chemical modelling of prestellar cores performed by \citet{b47}, however, found N$_{2}$H$^{+}$ column density increased with the contraction of the dense molecular cloud core.  \citet{b47} propose that due to its relatively large depletion time-scale  and time dependent column density, N$_{2}$H$^{+}$ may be a good indicator of core evolution. The increase in N$_{2}$H$^{+}$ column density during core collapse and its long depletion time-scale may explain the correlation seen in~\autoref{fig:molec_comparisons} a) between the peak brightness temperature of  HC$_{5}$N and N$_{2}$H$^{+}$, suggesting high levels of N$_{2}$H$^{+}$ occur in the same evolutionary phase that HC$_{5}$N is present. As N$_{2}$H$^{+}$ will likely outlast the fragile HC$_{5}$N, confined to a more limited evolutionary window, it thus may be possible to construct a chemical clock utilising these species. Peak brightness temperature of HC$_{5}$N has also been compared with the 1.2 mm infrared flux \citep{b40} in~\autoref{fig:molec_comparisons} b). Sources with HC$_{5}$N detections also appear in general to have greater 1.2 mm infrared flux than those with non-detections. \\

\setlength\fboxsep{0pt}
\setlength\fboxrule{0.5pt}

\subsection{Implications for star formation}
\noindent The results demonstrate that HC$_{5}$N may be present in the hot molecular core stage of high mass star formation, providing support for the \citet{b7} model. This model indicates that the column densities of HC$_{5}$N, HC$_{7}$N and HC$_{9}$N peak $\sim$1$\times$10$^{4.3}$ years after the onset of core collapse. The abundances fall off rapidly by $\sim$1$\times$10$^{5.3}$ years and then settle to much lower steady state abundances by $\sim$1$\times$10$^{6}$ years. This suggests the 35 detections are aged between 1$\times$10$^{4}$ and 1$\times$10$^{6}$ years \citep{b32}. Methanol masers (6.7 GHz) are expected to be present between 1$\times$10$^{4}$ and 4.5$\times$10$^{4}$ years, considerably narrowing the age estimate of the detections to within these times. It was also found that of the 35 detections, 13 were associated with OH masers and 20 with water masers. Of the non-detections nine were associated with OH masers and 13 with water masers. OH masers are expected between $\sim$2$\times$10$^{4}$ and $\sim$4.5$\times$10$^{4}$ years, while water masers are expected between $\sim$1.5$\times$10$^{4}$ and $\sim$4.5$\times$10$^{4}$ years \citep{b32}. \\

\citet{b7} propose that the column density ratios of HC$_{5}$N and CH$_{3}$CN,  which increases rapidly between 1$\times$10$^{4}$ and 1$\times$10$^{5.5}$ years, may be useful in establishing a chemical clock for the stages of star formation. As we have made detections of both these molecules simultaneously in numerous sources, we have shown this to be a plausible application of HC$_{5}$N detections to determine the age of hot molecular cores. \\

We have compared our observational results to the gas-phase \citet{b7} chemical model and identify HC$_{5}$N as a good candidate to construct a chemical clock. However, hot core chemistry may be a result of a complex combination of gas-phase and grain-surface chemical processes  (\citealt{b50}; \citealt{b51}). Grain-surface chemical models also show HC$_{5}$N may exist under the higher temperature conditions of hot cores (Garrod et al. 2008). The choice of model for comparison may, however, introduce some uncertainty in the approximate timeframes of the existence of this molecule and thus into a chemical clock constructed with HC$_{5}$N and other molecules such as CH$_{3}$CN or N$_{2}$H$^{+}$. Other uncertainties in such a chemical clock may be introduced by the dependence of the N$_{2}$H$^{+}$ abundance on that of CO and other species through which this ion can be depleted by proton transfer \citep{b52}. Despite this, HC$_{5}$N remains a good candidate with which to construct a chemical clock to further illuminate the evolutionary age of hot molecular cores and the progress of high mass star formation. \\

\section{Summary and Conclusions}
\noindent We have made 35 detections of HC$_{5}$N, and 44 non-detections toward methanol maser selected hot molecular cores at a 3$\sigma$ sensitivity limit of 54 mK. Previous modelling by \citet{b7} shows that HC$_{5}$N can form efficiently under the conditions of hot molecular cores and our observations support this. No clear relationships between detections of HC$_{5}$N with distance or associated methanol maser peak flux density were found. Sources with HC$_{5}$N detections also had detections of CH$_{3}$CN. In most HC$_{5}$N non-detection sources CH$_{3}$CN and N$_{2}$H$^{+}$ were also present. When HC$_{5}$N was detected, however, N$_{2}$H$^{+}$ was invariably brighter than when it was not detected. The results in general support the chemical modelling of \citet{b7}. However, contrary to preliminary results presented in that work, we find CH$_{3}$CN is detected in sources where detections and non-detections of HC$_{5}$N have been made. \\

The study of organic molecules in interstellar clouds is key to finding species which can be used as chemical clocks to help determine the evolutionary age of hot molecular cores and further understand the process of high mass star formation.  High resolution observations of the HC$_{5}$N J=14$\rightarrow$13 transition with the Atacama Large Millimeter Array (ALMA) in a larger sample of these cores would go far in furthering this research, providing imaging of the gas and their relative locations within the beams we have used in this study (the J=12$\rightarrow$11 transition at 31.95 GHz lies just outside Band 1 of ALMA, while the J=14$\rightarrow$13 at 37.28 GHz lies within it). This would allow verification of whether cold cores lie within the beam, or if a warm envelope surrounds the hot core. Single dish observations of HC$_{7}$N and HC$_{9}$N would also serve to verify whether long chain cyanopolyynes can exist under hot core conditions and provide a further test of the \citet{b7} model. \\

\section*{Acknowledgements}
\noindent 
We thank the anonymous referee for constructive comments which helped improve the paper. We gratefully acknowledge the use of MALT90 Survey spectral line data cubes in the course of this project. We also acknowledge Australian Research Council (ARC) support through Discovery Project DP0451893 awarded to The University of New South Wales and Macquarie University. Astrophysics at Queens University Belfast is supported by a grant from the Science and Technology Facilities Council (STFC). The Tidbinbilla 34 m DSS--34 Radio Telescope is part of the Canberra Deep Space Communication Complex, which is operated by CSIRO Astronomy and Space Science on behalf of NASA.\\

\label{lastpage}
\end{document}